\documentclass[twocolumn]{aastex631}

\newcommand{\ha}{H$\alpha$}

\begin{document}

\title[The Coolest Ae Stars]{Using the Coolest Ae Stars to Constrain Circumstellar Disk Viscosity}

\correspondingauthor{R.\ Anusha}
\email{araviku4@uwo.ca}

\author[0000-0002-9015-6417]{R.\ Anusha}
\affiliation{Department of Physics \& Astronomy, The University of Western Ontario \\ 
1151 Richmond Street, London, ON, Canada, N6A~3k7}

\author[0000-0002-0803-8615]{T.\ A.\ A.\ Sigut}
\affiliation{Department of Physics \& Astronomy, The University of Western Ontario \\ 
1151 Richmond Street, London, ON, Canada, N6A~3k7}
\affiliation{Institute for Earth \& Space Exploration (IESX), The University of Western Ontario \\  7134 Perth Drive, London, ON, Canada, N6A~5B7}

\begin{abstract}

Classical Ae (CAe) stars are main sequence, A-type stars with \ha\ emission but no signature of dust. They are thought to be the cool extension of the classical Be stars to lower masses. Recent surveys based on \ha\ spectroscopy have significantly increased the number of known CAe stars, with the population extending to spectral types as cool as A4 $(T_{\rm eff}\approx 8500\;$K). We compute the temperature structure of gaseous, circumstellar disks around A-type stars, including both radiative heating from the central star and viscous shear heating from the disk's rotation. We find that shear heating can become important for spectral types A2 and later and can act to increase the low temperatures predicted by purely radiatively heated disks. Our modeling indicates that the presence and strength of \ha\ emission for spectral types A2 and later significantly increases with the amount of shear heating included, and we propose that this dependence can be used to constrain the $\alpha$ viscosity parameter appropriate for CAe star disks.

\end{abstract}


\section{Introduction} 
\label{sec:intro}

Classical Ae (CAe) stars are main-sequence, A-type stars characterized by the presence of emission lines in their spectra, most notably that of \ha\ at $\lambda 6562.8\,$\AA. This emission provides evidence for a circumstellar disk \citep{Abt1973} which is believed to be an out-flowing (\textit{decretion}) disk. CAe stars are regarded as the cooler, later-type counterparts to the classical Be (CBe) stars \citep{Rivinius2013} and are observed to span spectral types A0 to A4 \citep{Anusha2021}. The CAe stars overlap with the A-type shell stars which show deep central absorption in \ha\ (but no emission) that falls below the expected photospheric profile; shell stars are thought to occur when the photosphere of a CAe star is directly obscured by the circumstellar disk due to the observer's viewing angle. The CAe definition explicitly excludes the Herbig Ae (HAe) stars, which are pre-main sequence objects still embedded within the remnants of their primordial accretion disks \citep{Waters1998}. 
HAe stars exhibit excess infrared (IR) emission due to thermal re-radiation from circumstellar dust, typically detected beyond 2$\mu$m \citep{Waters1998, Meeus2001, Vink2005}. In contrast, the disks associated with CAe stars are believed to originate from mass ejected by the rapidly rotating central A star \citep{Krticka2011a, Haubois2012,Kee2016}. These dust-free, gaseous decretion disks produce hydrogen emission lines through recombination in the gas \citep{Porter2003} and exhibit an infrared excess due to free-free and bound-free emission, similar to classical Be stars.
The number of known CAe stars is small, with fewer than 300 confirmed members \citep[see][and references therein]{Shridharan2021}. The limited number of CAe stars poses a challenge to comprehensively understanding this interesting class of objects, and highlights the need to probe processes that may distinguish them from their CBe counterparts. 

The first documented CAe/A-shell stars were 17~Lep and 14~Com \citep{Morgan1932}. \cite{Abt1973} studied CAe/A-shell stars in the optical and discovered eight from a sample of 35 fast-rotating, A-type stars. \citet{Abt1973} proposed that these objects represented the A-type equivalent of the CBe stars; however, \ha\ emission lines were not covered in this work. \cite{Andrillat1986} followed this with \ha\ observations for 20 CAe stars and found that the emission becomes weaker with later spectral subtypes, with the frequency of CAe stars declining rapidly towards A3 and disappearing after A4. Different selection criteria, such as anomalous IR emission \citep{Jaschek1991}, variability \citep{Irvine1979}, or fast rotation \citep{Ghosh1999}, have been employed to increase the likelihood of identifying CAe stars. \citet{Halbedel1994} studied the photometric variability of 41 CAe stars and reported no \ha\ emission beyond spectral type A4. The discovery of the nearly edge-on, dusty disk of $\beta$ \textit{Pictoris} (A0 IVe) \citep{Smith1984, Beust1990} sparked efforts to detect disks around main-sequence A-stars \citep{Walker1988,Oudmaijer1992} and to classify the source of emission. From various samples of known CAe/shell stars, \cite{Jaschek1986} cataloged 19 objects observed by IRAS and found that, similar to CBe stars, the \ha\ emission line strength of more than 50\% of CAe stars is variable with time. \cite{Bohlender2016} identified 30 new CAe/A-shell stars from medium resolution \ha\ spectra.

\cite{Zorec1997} collected CAe stars from diverse archival catalogs and suggested that $\approx\!2$\% of A0 stars and $\approx\!0.2$\% of A1-A2 stars have been classified as CAe stars. However, \cite{Monin2003} found that extrapolating the CBe frequency into the A-subtypes predicts that $\approx\!3$\% of A1 and A2 type stars should be CAe stars. This is almost 15 times higher than the estimate of \cite{Zorec1997}, highlighting the potential for either a large number of missing CAe stars or a sharp decline in CAe frequency at the A0 spectral type boundary. 

Recently, there have been extensive searches for CAe stars in large, spectroscopic data sets. \cite{Anusha2021} performed the first systematic study of CAe stars in the Galaxy using LAMOST \citep{Bai2021, cui2012} Data Release~5 (DR~5) and identified 159 new CAe stars. \cite{Shridharan2021} used a relaxed selection criteria to find an additional 74 CAe stars from LAMOST DR5. \cite{Hummerich2022} identified new A-shell stars via a search for strong Fe\,{\sc ii} multiplet~42 lines in B, A, F-type stars from LAMOST DR4. \cite{Zhang2022} compiled a catalog of 2754 early-type emission-line objects from LAMOST DR7, composed of CAe/CBe candidates, HAe/HBe stars, and nebular emission sources, which includes the 159 CAe stars considered in this current work. Although their study reports additional candidate CAe stars, they are not incorporated into the present study. \cite{Zhang2022} rely on spectral types assigned by the LAMOST pipeline which are subject to small but systematic errors, while \cite{Anusha2021} followed a rigorous selection criteria, ensuring a carefully curated sample, and conducted an independent re-evaluation of spectral types assigned to the CAe stars. To maintain consistency, and to avoid mixing samples classified with different criteria, our present study adopts the sample of 159 confirmed CAe stars and spectral classifications of \cite{Anusha2021}. 

The formation of decretion disks in CBe and CAe stars remains an active area of investigation. In CBe stars, rapid stellar rotation and non-radial pulsations are intrinsic characteristics \citep{Rivinius2013} and are widely accepted as key drivers of disk formation. Additional mechanisms have been proposed, though their applicability remains under debate and likely varies between individual systems. These include magnetic fields \citep{Neiner2005, Wade2014, Hubrig2017, Balona2021}, stellar winds \citep{Silaj2014, Kee2016}, and binary interactions \citep{Slettebak1982, Gray2009}. Notably, magnetic fields have not been conclusively detected in CBe stars, stellar winds, particularly in later-type B stars, are too weak to account for disk formation on their own, and not all CBe stars are confirmed members of close binary systems. While the precise mechanism that injects angular momentum into the inner edge of the disks is unknown, the subsequent evolution of the disks is well described by the transport of angular momentum due to gas viscosity. 

The viscous decretion disk (VDD) model of \citet{Lee1991} is very successful at reproducing the observed characteristics and time variability of the CBe stars \citep{Rivinius2013, Cure2022}. Almost all hydrodynamical modeling of CBe stars uses the $\alpha$ prescription for the viscosity ($\nu$) from \cite{Shakura1973a} with $\nu\equiv\alpha\,c_s\,H$, where $c_s$ is the gas sound speed, $H$ is the disk scale height, and $\alpha$ is the disk viscosity parameter, assumed to lie in the range $0\le\alpha\le1$. Estimates for the $\alpha$ value appropriate to CBe VDDs are estimated to lie in the range $0.1-0.3$ \citep{Bjorkman2005, Carciofi2012a, Klement2015} as determined by matching the viscous timescale, $t\sim R^2/\nu$, to observed variations in the disk emission. 
More recent modeling efforts examining long-term photometric monitoring of Be star disk evolution have revealed that the viscosity parameter $\alpha$ can, in some cases, significantly exceed the commonly cited range. For example, \cite{Rimulo2018} analyzed 81 outbursts from 54 Be stars and reported systematically higher values of $\alpha$ during disk formation phases ($\alpha \approx 0.63$), relative to dissipation phases ($\alpha \approx 0.29$). Similarly, detailed studies of the disk activity cycles of $\omega$ CMa by \cite{Ghoreyshi2018, Ghoreyshi2021} demonstrated that during outburst episodes, $\alpha$ can rise well above 0.3, with some phases approaching values as high as $\alpha \approx 1.0$. These findings suggest that the effective viscosity in Be star disks may be time-variable and dependent on the disk’s evolutionary state.

In addition to the transport of angular momentum in the disk, viscosity can lead to shear heating of the gas \citep{Pringle1981,Frank2002}. Shear heating must be present in the disks CBe and CAe stars due to their Keplerian rotation \citep{Lee1991}; however, for the CBe stars, shear heating is small compared to the radiative heating of the gas by the central star's photoionizing radiation field. However, as stellar effective temperatures drop below $T_{\rm eff}= 10^4\,$K for the A type stars, the direct radiative heating of the gas declines, and there is the possibility for shear heating to become more important. Emission lines in CAe stars occur as late as spectral type A4 ($T_{\rm eff}=8600,$K), and thermal modeling of CBe disks find that typical disk temperatures are $\approx\,60$\% of $T_{\rm eff}$ \citep{Millar1998,Carciofi2006,Sigut2009}. Thus, there is the open question of whether shear heating is actually required to produce \ha\ emission in the coolest objects.

The current paper focuses on the following two questions: (1) Can the observed \ha\ emission in Ae stars be explained by disks heated solely by photoionizing radiation from the central star or is additional heating due to viscosity required? (2) If shear heating is required, can this be used to estimate the disk viscosity $\alpha$ parameter appropriate for the CAe stars? The structure of this paper is as follows: Section \ref{sec:Calc} presents the calculations employing \texttt{Bedisk} and \texttt{Beray} code suite for the CAe stars, including shear heating (Section~\ref{sub:Shear}). Section~\ref{sec:Results} discusses the results of the disk temperature calculations and the impact of shear heating. Section~\ref{sec:ObsLAMOST} reviews the observed CAe sample of \citet{Anusha2021}, and Sections \ref{sub:Aefrac} and \ref{sub:CDFs} compare this sample to the theoretical disk modeling in an attempt to constrain the disk $\alpha$ parameter appropriate for CAe disks. Section~\ref{sec:Conclusions} presents conclusions and scope for future work.

\section{Calculations} 
\label{sec:Calc}

\subsection{The \protect\texttt{Bedisk} and \protect\texttt{Beray} Models}

Stellar parameters for the central A-type stars are given in Table~\ref{tab:astars}. The main sequence mass and $T_{\rm eff}$ calibrations are from Appendix~B of \citet{gray2022}, and the stellar radius was computed from the assumption that $\log(g)=4.00$ for all stars; the luminosity was then computed from the radius and $T_{\rm eff}$. Critical rotation velocities,\footnote{The rotation speed at which the effective gravitational acceleration at the equator vanishes.} set by the mass and radius \citep{Maeder2009book}, are also given in Table~\ref{tab:astars} and lie in the range $330 - 350\;\rm km\,s^{-1}$.

\begin{table}
\begin{center}
\caption{Adopted parameters for the central main sequence A stars.}
\label{tab:astars}
\begin{tabular}{crrcrrr}\hline\hline
Spectral & $M_*$ & $R_*$ & $L_*$ & $T_{\rm eff}$ & $\log(g)$ & $v_{\rm crit}$ \\
Type   & $(M_\odot)$ & $(R_\odot)$ & $(L_\odot)$ & (K) & & (km/s) \\ \hline
A0 & 2.46 & 2.60 & 51.4 & 9600 & 4.00 & 347 \\
A1 & 2.31 & 2.52 & 40.7 & 9200 & 4.00 & 342 \\
A2 & 2.21 & 2.46 & 35.7 & 9000 & 4.00 & 338 \\
A3 & 2.15 & 2.43 & 28.9 & 8600 & 4.00 & 336 \\
A4 & 2.10 & 2.40 & 25.7 & 8400 & 4.00 & 334 \\
A5 & 2.04 & 2.37 & 22.7 & 8200 & 4.00 & 331 \\ \hline
\end{tabular}
\end{center}
Notes.- The mass and $T_{\rm eff}$ (rounded to the nearest 200K) are from Appendix~B of \citet{gray2022}. The radius is computed assuming $\log(g)=4.00$, and $L_*=4\pi\,R_*^2\,\sigma\,T_{\rm eff}^4$.
\end{table}

Around each stellar model of Table~\ref{tab:astars}, axisymmetric, equatorial circumstellar disks were considered with gas densities defined by the parameters $(\rho_0,n,R_d)$ in the equation
\begin{equation}
\rho(R,Z)\equiv \rho_0 \,\left(\frac{R_*}{R}\right)^n\;e^{-(Z/H)^2} ~~~ \mbox{\rm for}~~ R_* \le R\le R_d \;.
\label{eq:rho}
\end{equation}
Here $(R,Z)$ are the cylindrical co-ordinates spanning the disk,\footnote{$R$ is the distance from the stellar rotation axis, and $Z$ is the height above ($Z>0$) or below ($Z<0$) the equatorial plane.} $R_*$ is the stellar radius from Table~\ref{tab:astars}, and $H$ is the disk scale height at each radial distance, defined as
\begin{equation}
\label{eq:Hscale}
\frac{H}{R} = \frac{c_s}{V_K} \;.
\end{equation}
Here $c_s$ is the gas sound speed, and $V_K$ is the Keplerian rotational speed at disk radius $R$. As $c_s\ll V_K$, the disk is geometrically thin. Equation~\ref{eq:Hscale} follows from the assumption of vertical hydrostatic equilibrium in the disk \citep{Sigut2009}. We fix the sound speed, $c_s$, at a temperature of 60\% of $T_{\rm eff}$ and in this case, the disk flares as $H=H_*\,(R/R_*)^{3/2}$ where $H_*$ is the scale height at $R=R_*$. This assumed disk temperature, 60\% of $T_{\rm eff}$, is {\it only used\/} to fix the disk scale height in this manner. The detailed temperature structure of the disk, $T(R,Z)$, is computed by enforcing energy balance in the disk as discussed below. Note that it is possible to compute models in which $c_s$ is computed consistently with the temperature structure of the disk \citep{Sigut2009}; however, the changes in the predicted \ha\ emission lines are small, and this approach is not pursued here.

The parameters $(\rho_0,n,R_d)$ in Equation~(\ref{eq:rho}) define specific circumstellar disks around the central stars of Table~\ref{tab:astars}. Following \citet{Sigut2023}, we have considered density $\rho_0$ in the range $10^{-12}$ to $2.5\cdot\,10^{-10}\;\rm g\,cm^{-3}$, power-law indices $n$ in the range of $1.5$ to $4.0$, and outer disk radii $R_d$ in range $5\,R_*$ to $50\,R_*$. This makes a total of 660 disk density models for each spectral type of Table~\ref{tab:astars}. These parameter ranges, as well as others defining the \ha\ calculations below, can be found in Table~\ref{tab:dparams}.

To compute the temperature structure of the disk, $T(R,Z)$, we used the \texttt{Bedisk} code \citep{Sigut2007,Sigut2018}. This code enforces radiative equilibrium in a gas of solar composition heated by the central star's photoionizing radiation field as represented by the photospheric emergent intensity, $I_{\mu\nu}(\tau_\nu=0)$ for $\mu\ge0$. These intensities were computed in LTE using the \texttt{Atlas9} stellar atmospheres code \citep{kurucz1991} for the star's $T_{\rm eff}$ and $\log(g)$ and were represented by 1221 frequency points in the range $50\,$\AA\ to $50\,\mu m$. As some disks considered in this work become quite cool ($T\!<\!5000\,$K), it is important that \texttt{Bedisk} includes the ten most abundant elements over several ionization stages \citep[see][]{Sigut2007}; therefore, abundant, low-ionization metals such as Fe\,{\sc i}, Na\,{\sc i}, Ca\,{\sc i}, and Mg\,{\sc i} are present to contribute to the gas electron density and heating and cooling rates when the excited hydrogen level populations become small. Molecule formation is not included in the models.

\subsection{Shear Heating}
\label{sub:Shear}

Due to the rapid decline of the central star's photoionizing radiation field with the decreasing $T_{\rm eff}$ of the A stars in Table~\ref{tab:astars}, we have added viscous shear heating to the calculation of the disk's thermal structure. The shear heating rate in $\rm ergs\,cm^{-2}\,s^{-1}$ is
\begin{equation}
D(R)=\frac{1}{2}\nu\Sigma\,\left(R\frac{d\Omega}{dR}\right)^2 \;,
\end{equation}
\citep{Pringle1981,Frank2002} where $\nu$ is the gas viscosity, $\Sigma$ is the disk surface density, and $d\Omega/dR$ is the disk angular velocity gradient. We write this as
\begin{equation}
D(R) = \int_{-\infty}^{+\infty}\,\frac{1}{2}\nu\,\rho\,\left(R\frac{d\Omega}{dR}\right)^2\,dZ \equiv \int_{-\infty}^{+\infty}\,d(R,z) \,dZ \;,
\end{equation}
where we identify the integrand as the volumetric heating rate, $d(R,Z)$, in $\rm ergs\,cm^{-3}\,s^{-1}$. Using the $\alpha$-prescription for the viscosity, $\nu=\alpha\,c_s\,H$ \citep{Shakura1973a} with $\alpha$ a constant parameter, the sound speed, $c_s^2=\gamma P/\rho$, the assumed Keplerian rotation for the disk, $\Omega=\sqrt{GM/R^3}$, and the definition of the disk scale height $H$ above, we find
\begin{equation}
D(R) = \int_{-\infty}^{+\infty}\, \left(\frac{9}{8}\gamma\,\alpha\,P\,\Omega\right)\,dZ \,.
\end{equation}
We choose $(9/8)\gamma = 3/2$ for consistency\footnote{This gives $\gamma=1.333$ which lies between isothermal $(\gamma=1)$ and adiabatic ($\gamma=5/3$) motion for the turbulent eddies that underlie the viscosity.} with Equation~(7) of \citet{Lee1991}, giving a volumetric heating rate of
\begin{equation}
\label{eq:shearheat}
d(R,Z) = \frac{3}{2}\alpha\,P\,\Omega \,,
\end{equation}
which has been incorporated into \texttt{Bedisk} to represent shear heating. Note that the magnitude of the heating rate is set by the value of the $\alpha$ parameter in the viscosity, and we consider values $\alpha=0.01$, $0.1$, $0.3$, and $1.0$. Current estimates of $\alpha$ in CBe stars disks, all based on estimates of the viscous timescale $t\sim R^2/\nu$, are in the range $0.1$ to $0.3$, although will considerable uncertainty\footnote{While we have assumed $\alpha$ to be constant over the entire disk, variations of $\alpha$ with $R$ and/or $Z$ are certainly possible \citep[see also][]{Granada2021}.}\citep{Granada2021}. Thus, the \texttt{Bedisk} star+disk models are fully defined by the parameters $(\mbox{\rm Spectral Type};\rho_0,n,R_D;\alpha)$, making a total of 2640 models for each spectral type of Table~\ref{tab:astars}.

\subsection{H\protect{$\alpha$} Calculations}
\label{sub:Hacalcs}

\begin{table}
\caption{Ranges for the disk parameters used in the \ha\ calculations.}
\label{tab:dparams}
\begin{center}
\begin{tabular}{ll}\hline\hline
Parameter     &  Values \\\hline
Spectral Type & (6): A0, A1, A2, A3, A4, A5 \\
$\log\rho_0$  & (15): $-12.0$ to $-9.60$ with $\Delta\log\rho_0=0.17$ \\
$n$           & (11): $1.50$ to $4.00$ with $\Delta n=0.25$ \\
$R_d/R_*$      & (4): $5$, $15$, $25$, $50$ \\
$\alpha$      & (4): $0.01$, $0.10$, $0.30$, $1.00$ \\
$i$           & (11): $0$ to $90^\circ$ with $\Delta i = 10$ plus $i=85^\circ$\\\hline
\end{tabular}
\end{center}
Notes.- All parameter combinations total 174,240 individual \ha\ profiles.
\end{table}

Corresponding to each model defined by the star+disk parameters, we have computed the corresponding \ha\ profile. This is done to ensure that the models of focus are the ones that produce detectable \ha\ emission and would be classified as CAe stars; this point is further discussed below. To compute the \ha\ profiles, the \texttt{Beray} code was used \citep{Sigut2018}. Here the equation of radiative transfer is solved along a large number of parallel rays directed at a distant observer as defined by the inclination angle $i$, the angle between the stellar rotation axis and the observer's line-of-sight. An inclination of $i=0^\circ$ represents a pole-on star/face-on disk, while $i=90^\circ$ represents an equator-on star/edge-on disk. Systems with $i\ge\!80^\circ$ often exhibit deep central absorption features in \ha, observationally recognized as the A-shell stars previously discussed. For rays that terminate on the stellar surface, an appropriately Doppler-shifted, LTE, photospheric, absorption \ha\ profile (corresponding to the $T_{\rm eff}$ and $\log(g)$ of the central star) is used as the upwind boundary condition for the transfer equation. Rays that pass through the disk but do not terminate on the star assume a zero intensity boundary condition. This procedure naturally produces a \ha\ profile for the star+disk system that can be directly compared to observations; for example, in the limit $\rho_0\rightarrow 0$ in Eq.~(\ref{eq:rho}), \texttt{Beray} will yield the \ha\ profile of the star alone. We have computed \ha\ profiles for 11 inclination angles, $i=0,10,20,\ldots,70,80,85,90^\circ$, with the angle $i=85^\circ$ added to better sample the development of shell absorption. Thus, for each spectral type of Table~\ref{tab:astars}, there are 29,040 individual \ha\ profiles. The total number of \ha\ profiles computed was 174,240 (Table~\ref{tab:dparams}), reflecting the six spectral types of Table~\ref{tab:astars}. An individual \ha\ profile is fully defined by the parameters $(\mbox{\rm Spectral Type};\rho_0,n,R_D;\alpha;i)$.

All computed \ha\ profiles were automatically classified as to whether or not they exhibited emission. Each computed \ha\ profile was convolved down to a spectral resolution of ${\cal R}=2000$\footnote{This resolution is chosen to closely match the observed sample discussed in Section~\ref{sec:ObsLAMOST}.}, and this smooth profile was numerically differentiated to determine local minima and maxima. An integer encodes the pattern of minima (with 1) and maxima (with 2). For example, the classification `1' indicates a single minimum or a pure absorption profile with no emission, and a classification of `121' indicates a single emission peak in the center of the photospheric absorption \ha\ line. In practice, almost all profiles fit into the categories of either a pure absorption profile (class 1) or a singly or doubly-peaked emission component in the center of the photospheric absorption H$\alpha$ line (classes 121 or 12121 respectively). Thus, the class 1 profiles flag models $(\mbox{\rm Spectral Type};\rho_0,n,R_D;\alpha;i)$ that predict no \ha\ emission and thus do not satisfy the observational criteria for classification as a CAe star.

All calculations were performed assuming a stellar equatorial rotation speed of 80\% of the critical value listed in Table~\ref{tab:astars}. While rapid stellar rotation is well known to result in gravitational darkening in which the stellar surface becomes rotationally distorted and the $T_{\rm eff}$ and $\log(g)$ become latitude dependent \citep{vonZeipel1924,Collins1965,Espinosa2011}, we have not included this effect in these exploratory calculations for the CAe stars, although \texttt{Bedisk} has the capability to include gravitational darkening \citep{McGill2011}. The effect of gravitational darkening on the thermal structure of Be star disks is relatively small \citep{McGill2011,McGill2013}, and these calculations will be left to a future work.

\section{Results} 
\label{sec:Results}

\subsection{Disk Temperatures} 
\label{sub:Disktemps}

Given the star+disk parameters, \texttt{Bedisk} computes the temperature structure, $T(R,Z)$, throughout the disk. To compare disk temperatures over the large number of models computed, it is useful to associate a single temperature measure to each model, and the density-weighted, average disk temperature,
\begin{equation}
\label{eq:trho}
<\!T_{\rho}\!> \equiv \frac{2\pi}{M_D}\,\int_{-\infty}^{+\infty}\,dZ\,\int_{R_*}^{R_D}\,R\,dR\;\left\{\rho(R,Z)\,T(R,Z)\right\}\;,
\end{equation}
was chosen. In this equation, the gas density $\rho(R,Z)$ is from Equation~(\ref{eq:rho}), $M_D$ is the total mass of the disk following from $\rho$, and $T(R,Z)$ are the radiative equilibrium temperatures found by \texttt{Bedisk}.

\begin{figure*}
\centering
\includegraphics[width=1.0\textwidth]{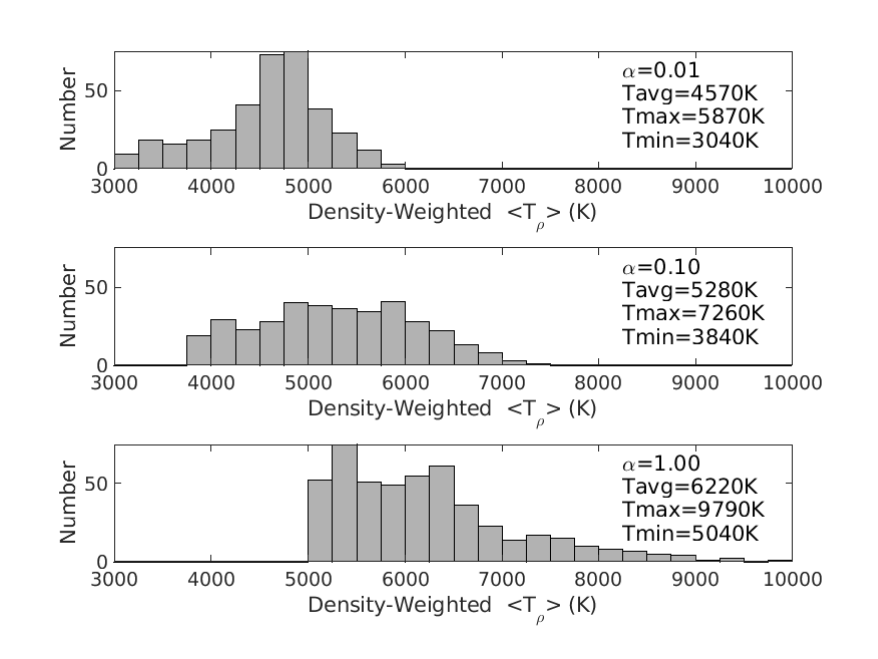}
\caption{Histograms of the density-weighted disk temperatures (Eq.~\ref{eq:trho}) for all spectral types of Table~\ref{tab:astars} and disk parameters of Table~\ref{tab:dparams} resulting in detectable \ha\ emission. The panels are for different amounts of shear heating, controlled by the $\alpha$ parameter: $\alpha=0.01$ (top), $\alpha=0.10$ (middle), and $\alpha=1.00$ (bottom).}
\label{fig:Thistograms}
\end{figure*}

Figure~\ref{fig:Thistograms} shows histograms of $<\!T_{\rho}\!>$ for three different levels of shear heating corresponding to $\alpha=0.01$, $0.10$, and $1.00$ for all spectral types of Table~\ref{tab:astars}. For each choice of $\alpha$, the histogram is composed of only those star+disk models which exhibited \ha\ emission. To classify the models (as described above), the \texttt{Beray} calculations for $i=60^\circ$ and $R_d=50\,R_*$ were used. From the figure, it is clear that shear heating can have a significant impact on average disk temperatures. For $\alpha=0.01$, essentially zero shear heating, temperatures as low as $3000\,$K are realized, with an average $<\!T_{\rho}\!>$ over the A0-A5 spectral types of about $4600\,$K.\footnote{It might seem contradictory that such cool disks could produce {\it any \ha\ emission at all}. However, $<\!T_{\rho}\!>$ is just an average for the disk. In reality, there are usually strong temperature variations in photoionized disks, particularly in the vertical ($Z$) direction. Such disks typically have hot sheaths above and below the equatorial plane that are directly illuminated by the star, with cooler regions near the equatorial plane where the star's radiation suffers significant extinction due to high optical depths along all rays back to the star-- see, for example, \cite{Sigut2007}.} As the shear heating is increased, the disk temperatures steadily increase with disks cooler than $<\!T_{\rho}\!>=5000\,$K essentially disappearing for $\alpha=1.00$.

\begin{figure*}
\centering
\includegraphics[width=1.0\textwidth]{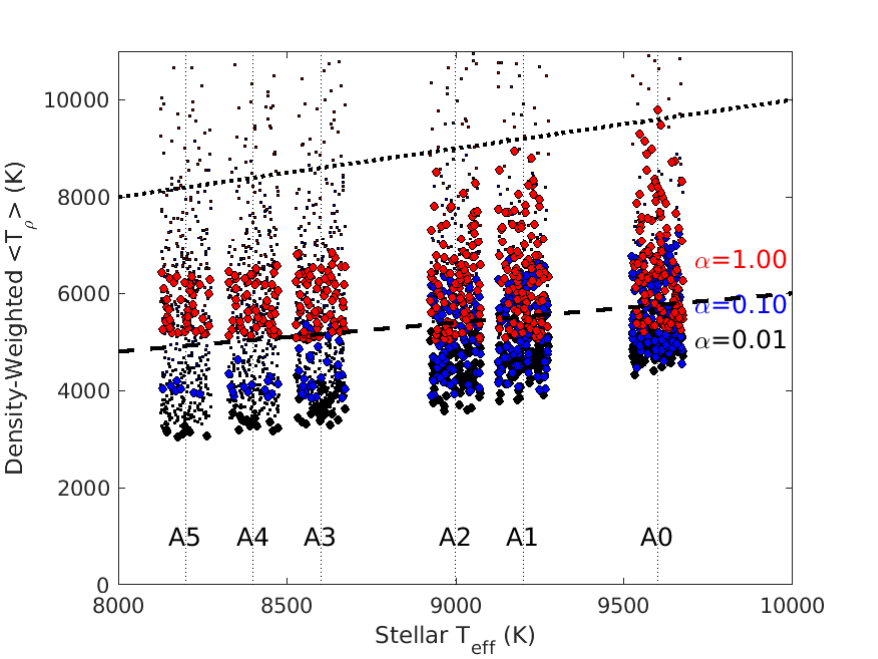}
\caption{The density-weighted disk temperature (Eq. 7) as a function of the central star’s effective temperature. Each point represents $\langle T_{\rho} \rangle$ (Eq. 7) for a different combination of parameters (spectral type, $n$, $\rho_0$, $\alpha$), with the corresponding $T_{\rm eff}$ values randomly jittered by ±100 K for visibility. Symbol colors indicate different levels of shear heating governed by the viscosity parameter $\alpha$: $\alpha$ = 0.01 (black), $\alpha$ = 0.10 (blue), and $\alpha$ = 1.00 (red). Large symbols correspond to models that produce detectable H$\alpha$ emission; small symbols denote non-emitting models. The lines $\langle T_{\rho} \rangle = T_{\rm eff}$ (thick dotted) and $\langle T_{\rho} \rangle = 0.6~T_{\rm eff}$ (thick dashed) have been added for reference. Faint, vertical dotted lines mark the $T_{\rm eff}$ values corresponding to the spectral types listed in Table 1.}
\label{fig:TdiskTeff}
\end{figure*}

Figure~\ref{fig:TdiskTeff} plots $<\!T_{\rho}\!>$ versus the central stellar $T_{\rm eff}$, showing the behavior at different spectral types. Unlike the previous figure, Figure~\ref{fig:TdiskTeff} shows $<\!T_{\rho}\!>$ for {\it all computed models}, not just those that show \ha\ emission. Models that satisfy the requirement of \ha\ emission are shown as large symbols (color-coded by the choice of $\alpha$), whereas models that do not show \ha\ emission, and would not be classified as CAe stars, are shown as small dots. There are several points of interest in this figure. Firstly, the segregation of models by the amount of shear heating (controlled by the $\alpha$ parameter) becomes noticeable around spectral type A2, and for cooler stars, the temperatures of all $\alpha=1.00$ disks with detectable \ha\ emission lie above the $<\!T_{\rho}\!>=0.6\,T_{\rm eff}$ line. The $<\!T_{\rho}\!>=5000\,$K plateau seen in Figure~\ref{fig:Thistograms} is clearly visible (red symbols). Secondly, the fraction of models that produce detectable \ha\ emission (compare large symbols with small ones) is strongly dependent on the $\alpha$ parameter for spectral types cooler than A2 (this is further discussed below). Thirdly, some high $<\!T_{\rho}\!>$ values are present, some of which {\it exceed\/} the stellar $T_{\rm eff}$ when shear heating is present. It is important to note, however, that such disk temperatures appear almost exclusively for models that do not exhibit \ha\ emission and would not be classified as CAe stars.

Models that do not exhibit detectable \ha\ emission are too rarefied to have significant \ha\ emissivity, i.e.\ they have a combination of small $\rho_0$ and/or large $n$ in Equation~(\ref{eq:rho}) that leads to very low $\rho(R,Z)$ throughout the entire disk. The gas cooling processes in the disk, namely the escape of collisionally-excited line radiation, the escape of photons formed by recombination, and free-free emission, all vary with $\rho^2$. Photoionization heating, while proportional to $\rho$, also requires a photon in the initial state, and the radiation energy density decreases with geometric dilution as one moves further away from the star. Shear heating, on the other hand, is proportional to $\rho$. For models in which shear heating becomes important, the ratio of heating-to-cooling scales as $1/\rho$, and thus low density regions have difficulty cooling, driving higher temperatures. This can dominate the temperatures in rarefied disks with small $\rho_0$ and large $n$. However, these low density disks have low emissivity and thus make little contribution to an observable diagnostic such as emission in \ha. This is the reason why we have been careful to classify all models as to whether or not they can produce emission in \ha\ and hence could be observed as CAe stars.

It is not clear to what extent the high disk temperatures discussed above are an artifact of assuming a constant $\alpha$ parameter for the entire disk volume. While molecular viscosity is independent of the gas density,\footnote{A fact that surprised Maxwell when he derived this result in 1860 \citep{Vincenti1965}.} molecular viscosity is incapable by many orders of magnitude to make viscous accretion or decretion disks viable \citep{Frank2002}; the viscosity in CAe and CBe stars is likely driven by turbulence and/or magnetic fields which could introduce some dependence on location in the disk \citep{Desch1998}. While this is an interesting point, we did not pursue models with ad-hoc parametrizations of $\alpha$, i.e.\ $\alpha(R,Z)$, as the disks affected did not present \ha\ emission and are not relevant to the observed samples of CAe stars discussed here. 

\begin{figure}
\centering
\includegraphics[width=1.0\columnwidth]{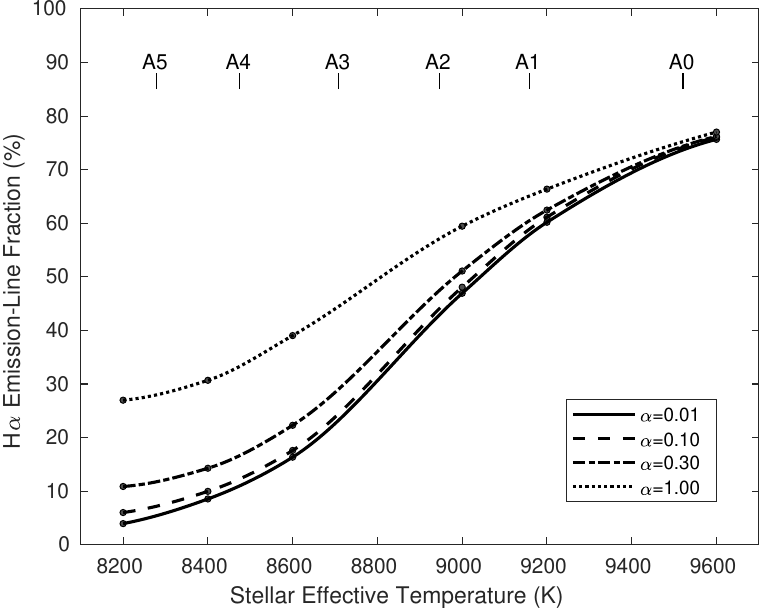}
\caption{The fraction of computed models (Table~\ref{tab:dparams}) that result in detectable H$\alpha$ emission as a function of the central star's $T_{\rm eff}$. Four lines are shown corresponding to the choices for the $\alpha$ parameter setting the viscosity: $\alpha=0.0$ (solid), $\alpha=0.1$ (dashed), $\alpha=0.3$ (dash-dot), and $\alpha=1.0$ (dotted). The circles are the calculations, and the lines are spline fits to this data. The spectral type - $T_{\rm eff}$ calibration is from \citet{gray2022} and is not rounded to the nearest 200~K as in Table~\ref{tab:astars}.}
\label{fig:emission}
\end{figure}

\begin{figure}
\centering
\includegraphics[width=1.0\columnwidth]{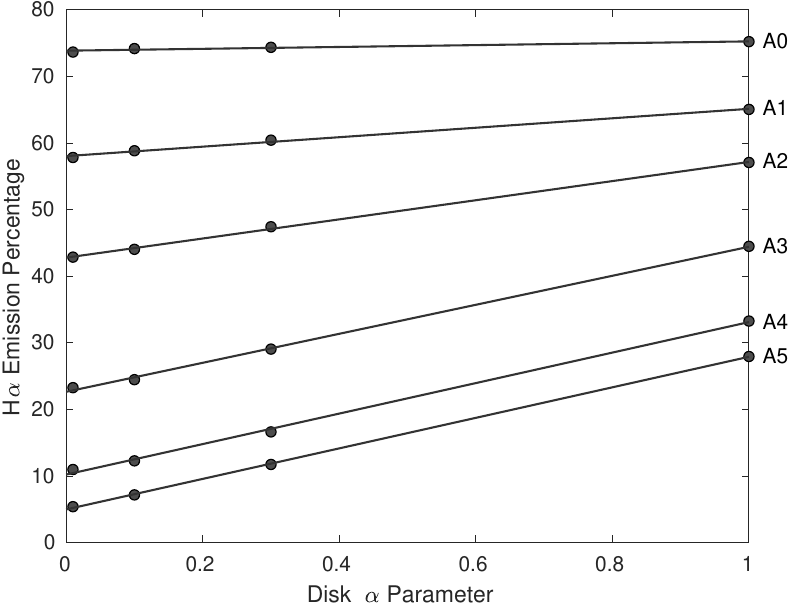}
\caption{The dependence of the \ha\ emission fraction on the disk $\alpha$ parameter controlling the amount of shear heating. The $\alpha$ value associated with each line is given in the legend. The circles are the calculations (Figure~\ref{fig:emission}) and the lines are linear fits to the data. The spectral type identifying each line is to the right.}
\label{fig:emission_alpha}
\end{figure}

Figure~\ref{fig:emission} shows for each $T_{\rm eff}$ the percentage of star+disk models (of the 7260 computed for each spectral type and $\alpha$ combination) that result in detectable \ha\ emission for the various levels of shear heating considered as controlled by the $\alpha$ parameter. Note that $\alpha=0.30$ is also included in this figure. Figure~\ref{fig:emission_alpha} shows how the \ha\ emission fraction at each spectral type depends on $\alpha$. For the earliest spectral type, A0, there is essentially no dependence of the percentage of computed models with \ha\ emission on $\alpha$. However, this changes at spectral type A2 where shear heating can become important. By spectral type A4, there is a strong dependence of the emission fraction on $\alpha$ in the sense that larger $\alpha$ values lead to a larger fraction of models that exhibit emission; the dependence on $\alpha$ is essentially linear at each spectral type as illustrated in Figure~\ref{fig:emission_alpha}. In both of these figures, the absolute fraction of models with emission is somewhat arbitrary as the set of star+disk models considered is somewhat arbitrary; for example, we could have instead retained only models that give emission at A0 and then kept this set fixed for the later spectral types, shifting the curves upward. However, the important point is that the set of models is kept the same for all spectral types so that relative changes are significant. Figures~\ref{fig:emission} and \ref{fig:emission_alpha} strongly suggest that the fraction of Ae stars detected at the latest spectral types (A3 and A4, and potentially A5 if there are such objects) might be able to be used to constrain $\alpha$ through its effect on disk temperatures and \ha\ emission. We look to this point in the next section.

\section{Comparison with the LAMOST-DR5 A\lowercase{e} Star Sample}
\label{sec:ObsLAMOST}

In this section, we investigate the possibility of using the Ae star sample of \cite{Anusha2021} to constrain the disk $\alpha$ viscosity parameter appropriate to CAe star disks using both the observed variation of the fraction of detected Ae stars with spectral type and the \ha\ equivalent width distributions for different spectral types.

\begin{figure*}
\centering
\includegraphics[width=1.0\textwidth]{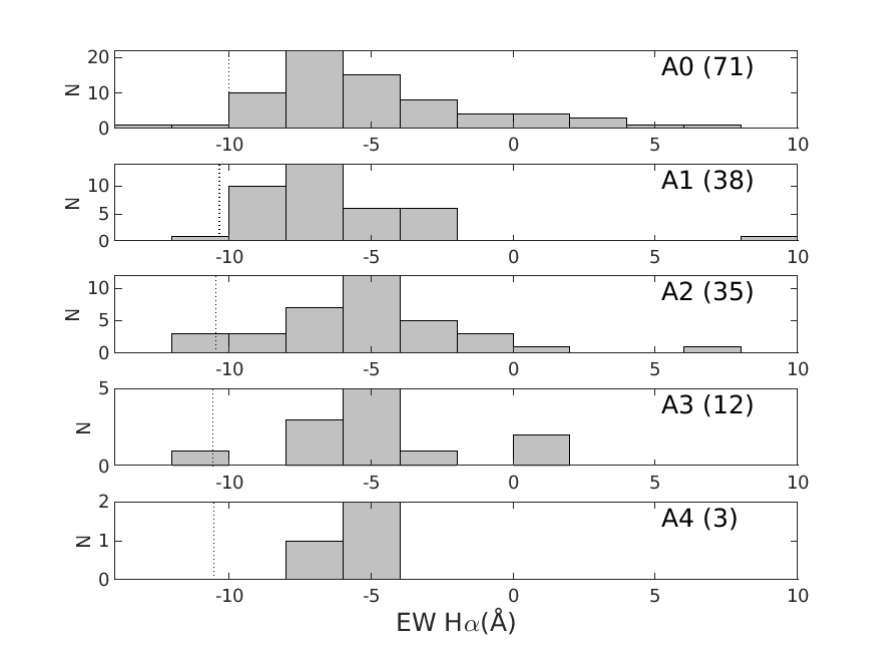}
\caption{Histograms of the measured LAMOST \ha\ equivalent widths (in \AA ngstroms) for each spectral type in the Ae star sample of \citet{Anusha2021}. The number of sample stars for each spectral type is given in brackets. EW$<0$ is net absorption, whereas EW$>0$ is net emission. The dotted line in each panel gives the LTE, photospheric absorption \ha\ equivalent width for $\log(g)=4.00$ at that spectral type.}
\label{fig:ewhist}
\end{figure*}

The CAe sample from \cite{Anusha2021} is based on the data release 5 (DR5) survey of the Large Sky Area Multi-Object Fiber Spectroscopic Telescope (LAMOST). LAMOST is operated by the Chinese Academy of Sciences \citep{cui2012}, and they survey was initiated by the National Astronomical Observatories of China (NAOC). \cite{Anusha2021} utilized the stellar spectra from LAMOST DR5 to identify and characterize CAe stars based on their \ha\ emission, IR color, and IR excess. This work identified 159 CAe stars which are utilized for the present study. Additionally, we cross-matched these targets with the LAMOST low and medium resolution spectral catalogs from data release 8 (DR8), discovering 23 new observations of these stars. This makes a total of 220 potential \ha\ spectra for the 159 CAe stars from \citet{Anusha2021}. For stars with multiple spectra available, we included only the single spectrum with the highest S/N in the vicinity of \ha. The mean distances of these targets, cross-matched with \citet{Bailer-Jones2021}, ranges from 350~pc to 4900~pc, with errors increasing to up to 15\% for distances greater than 3000~pc. The V-magnitudes of the targets, cross-matched with APASS~DR9 \citep{Henden2015}, range from 10.0 to 16.8 mag. All targets are positioned towards the Galactic anti-center, with more than 95\% of the targets located between $120^\circ\leq\ell\leq 240^\circ$ as the LAMOST survey focuses on the Galactic anti-center.

The spectral types of the 159 CAe star sample range from A0 to A4, with the breakdown: A0 (71 objects), A1 (38 objects), A2 (35 objects), A3 (12 objects), and A4 (3) objects. No A5 or later stars were identified with \ha\ emission. The spectral type of each LAMOST DR5 spectrum is estimated using the LAMOST Stellar Parameter Pipeline \citep[LASP;][]{Wu2014}. however, the presence of a potentially veiled continuum and the presence of emission lines for CAe stars can complicate the spectral classification \citep{Doazan1991, Hummel2000}. Therefore, \citet{Anusha2021} re-estimated the spectral types semi-automatically using a template matching technique with the MILES stellar spectral library \citep{Sanchez2006}. 

For each object, the \ha\ line was extracted and continuum normalized (at $\pm 37$\AA) in order to measure it's equivalent width (EW). In contrast to \citet{Anusha2021}, a photospheric \ha\ EW appropriate to the star's spectral type {\it was not subtracted} from the measured EW. We adopt the convention that $EW>0$ represents net emission and $EW<0$, net absorption. The average S/N around the \ha\ lines is typically $\approx\!90$, and the re-normalized spectra yielded \ha\ EWs ranging from $-12.1\,$\AA\ to $+24.2\,$\AA. The lines were also morphologically classified into the categories defined in Section~\ref{sub:Hacalcs}. From the sample of 159 Ae stars, 153 (96\%) were classified with an index of 121 (single emission peak in an absorption trough), two objects with an index of 12121 (doubly-peaked emission in an absorption trough), and four objects with an index of 2 (single emission peak with no detectable absorption).

Figure~\ref{fig:ewhist} shows the distributions of measured \ha\ equivalent widths as a function of spectral type. The steep decline in detected Ae stars for later spectral types, A3 and A4, is apparent. With \ha\ distributions for five spectral types, we performed ten, two-sample Kolmogorov-Smirnov (KS) tests, one for each pair of spectral types, to see if the distribution pairs are consistent with the same underlying distribution. The KS test is used because an unbinned test is desirable given the small number of objects with A3 and A4 spectral types. The results are presented in Table~\ref{tab:ksew} which gives, for each pair of spectral types, the $\log\!10$ of the probability ($\log P$) that a KS statistic equal to or larger than observed is consistent with random sampling of the same underlying distribution. At the 1\% level, no comparisons fail. At the 5\% level ($\log P=-1.30$) only the A1-A2 comparison fails. The interpretation of this failure is not entirely clear. Given the results of the calculations in the previous section, one might have expected failures between earlier (A0 and A1) and later (A3 and A4) spectral types; however, the very small number of objects at A3 and A4 precludes sensitive tests. 

\begin{table}
\caption{Two-sample KS tests for each pair of equivalent width distributions shown in Figure~\ref{fig:ewhist}. Each entry gives the $\log-10$ of the probability that the equivalent width distributions of the spectral types in the row and column headers are drawn from the same underlying distribution.}
\label{tab:ksew}
\begin{center}
\begin{tabular}{c|ccccc}\hline\hline
 Spectral Type   &   A0    &   A1    &   A2     &  A3     &  A4 \\\hline
 A0 &  0.000 &  -0.991  & -0.124  & -0.123  & -0.112 \\
 A1 &        & 0.000    &  -1.645 &  -0.975 & -0.260 \\
 A2 &        &          &   0.000 &  -0.176 & -0.025 \\
 A3 &        &          &         &  0.000  & -0.051 \\
 A4 &        &          &         &         &  0.000 \\ \hline
\end{tabular}
\end{center}
\end{table}

\begin{figure}
\centering
\includegraphics[width=1.0\columnwidth]{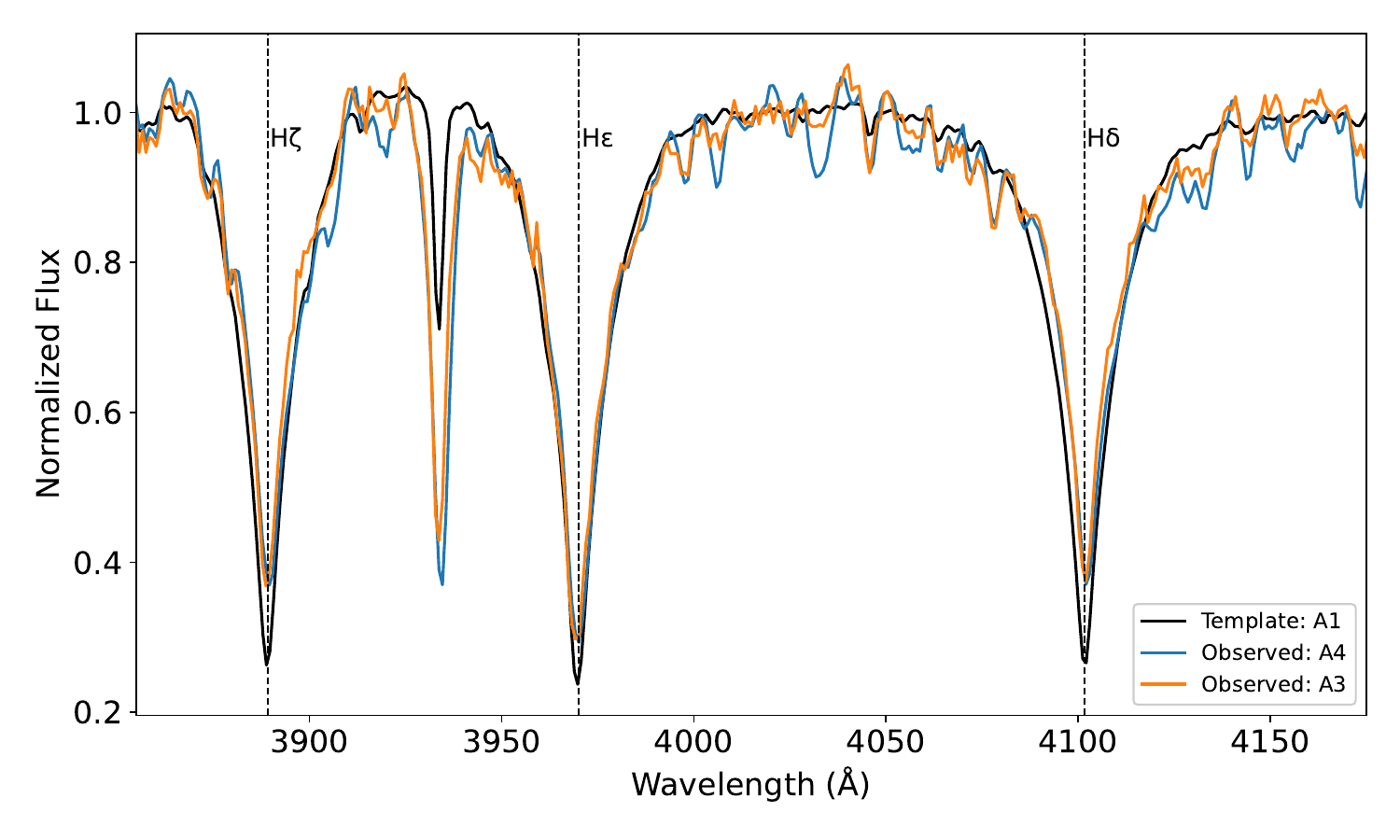}
\hfill
\caption{Comparison of the optical spectrum of a LAMOST A4-type CAe star (blue line, LAMOST J114805.60+412843.2) and A3-type CAe star (orange line, LAMOST J184640.36+425427.9) with the A1-type MILES template (black line). Note the mismatches in the central depths of the hydrogen series H{$\zeta$}, H{$\epsilon$}, and H{$\delta$}, as well as in the strength of the Ca{\sc ii} K line at $\lambda3934\,$\AA.}
\label{fig:template}
\end{figure}

The small number of objects of spectral types A3 and A4 does make the analysis to follow sensitive to the possibility of spectral type misclassification, i.e.\ these objects might be misclassified earlier spectral types. For this reason, we have visually re-compared the spectra for the A3 and A4 objects with the available MILES templates. Figure~\ref{fig:template} compares two of the CAe stars (spectral type A4: LAMOST J114805.60+412843.2, and spectral type A3: LAMOST J184640.36+425427.9) to the A1V template in the vicinity of $4000\,$\AA. Noteworthy is the Ca\,{\sc ii} K line at $\lambda3934\,$\AA. This line is very temperature sensitive over the range of the A stars, with its strength increasing with decreasing $T_{\rm eff}$. As can be seen from the figure, the Ca\,{\sc ii} K line in the A4 and A3 spectrum is much stronger than in the A1 template. The other comparisons gave similar results, and there is no clear evidence that the A3 or A4 spectral types are misclassifications.

\subsection{Fraction of Ae Stars with Spectral Type}
\label{sub:Aefrac}

In this section, we will test if our CAe star \ha\ calculations above are consistent with the steep decline in the number of CAe stars with spectral type detected by the LAMOST sample of \citet{Anusha2021}. 

Consider a sample of $N$ stars. The number of stars of class c1 in this sample that have detectable \ha\ disk emission can be written as
\begin{equation}
N^{e}_{c1}=N\,f_{c1}\,D_{c1}\,E_{c1} \,.
\end{equation}
Here $f_{c1}$ is the fraction of sample stars that are of class c1, $D_{c1}$ is the fraction of such stars that have disks, and $E_{c1}$ is the fraction of disks that result in detectable \ha\ emission. Given some other class c2, the ratio of the number of stars with detectable \ha\ disk emission in class c2 to those in class c1 is
\begin{equation}
\frac{N^e_{c2}}{N^e_{c1}} = \frac{f_{c2}}{f_{c1}}\,\frac{D_{c2}}{D_{c1}}\,\frac{E_{c2}}{E_{c1}}\,.
\label{eq:counts}
\end{equation}
In this expression, the ratio $f_{c2}/f_{c1}$ can be obtained directly from the survey. The ratio $D_{c2}/D_{c1}$ could be in general complex, but in the present case of A-type main sequence stars over a limited range of adjacent spectral types, it does not seen unjustified to {\it assume that\/} $D_{c2}/D_{c1}\approx 1$. The third ratio, $E_{c2}/E_{c1}$, can be deduced from Figure~\ref{fig:emission_alpha}. To apply this to the LAMOST sample, we take class c2 to be spectral types A3 and A4, and class c1 to be spectral types A0 and A1; we have combined spectral types in this manner for several reasons. Firstly, combining A3 and A4 increases the number of objects in the cool bin. Secondly, the reason for the large difference in the number of A0 and A1 stars (over a factor of two) is not entirely clear to us\footnote{This may have to do with the availability (or lack thereof) of A spectral subtype templates used in the classifications.}. Finally, these two combined bins straddle where the steep decline starts in \ha\ emission numbers. 

From the spectral type counts of the LAMOST-DR5 sample, $N({\rm A34})=76087$ and $N({\rm A01})=89036$; therefore, $f_{A34}/f_{A01}\approx 0.86$\footnote{Based on the Salpeter IMF and crudely assigning $c1$ (A0-A1) to mass bin $[2.30,2.50]\,M_\odot$ and $c2$ (A3-A4) to mass bin $[2.00,2.20]\,M_\odot$ solar masses, one might have expected the ratio $f_{c2}/f_{c1}\approx 1.4$.}. \citet{Anusha2021} find $N^e_{A34}/N^e_{A01}=15/109$ (Figure~\ref{fig:ewhist}). Figure~\ref{fig:emission_alpha} suggests $E_{A01}\approx 0.7$. Putting this together, we can solve Eq.~(\ref{eq:counts}) for $E_{A34}$ to find
\begin{eqnarray}
E_{A34} & = & \left(\frac{f_{A01}}{f_{A34}}\right)_{\rm obs}
\left(\frac{N^e_{A34}}{N^e_{A01}}\right)_{\rm obs} 
\left(E_{A01}\right)_{\rm fig4}\nonumber\\ 
        &\approx  & 1.2 \left(\frac{15}{109}\right)\,0.7\;,
\end{eqnarray}

or $E_{A34}\approx 12\%$ which is consistent with Figure~\ref{fig:emission_alpha} with $\alpha\le0.1$. This agreement might be somewhat fortuitous given the uncertainties. If we assume the error is dominated by the $\sqrt{N}$ uncertainty in the A3/A4 counts, the $1\sigma$ range of $E_{A34}$ is approximately 8\% to 15\%; therefore, the result is not statistically inconsistent at $2\sigma$ with $\alpha\le0.30$. In addition, this result is based on the assumption that $D_{c2}/D_{c1}\approx 1$ (that similar disks occur for both early- and late-type CAe stars) which might not be true. Nevertheless, our \ha\ calculations seem consistent with the steep decline in CAe star numbers at essentially the correct A2 spectral type.

\subsection{Modeling The LAMOST Equivalent Width Distributions}
\label{sub:CDFs}

A stricter comparison with the CAe star sample of \cite{Anusha2021} is to compare the range of \ha\ emission strengths, as opposed to the binary ``Yes/No" emission classification of the previous section using CAe star counts. In this section, we attempt to match the observed \ha\ equivalent width distributions for each spectral type shown in Figure~\ref{fig:ewhist}. In practical terms, the cumulative distribution (CDF) of the observed \ha\ equivalent widths will be compared with samples constructed from the models of Section~\ref{sub:Hacalcs} using a two-sample KS test in order to avoid binning the observed data.

We use the \ha\ calculations of Section~\ref{sec:Calc} to build CAe samples of 100 stars for each spectral type and viscous $\alpha$ value that can then be compared to the observed samples in Figure~\ref{fig:ewhist}. To do this, we need to specify probability distributions for the disk density parameters $(\rho_0,n,R_d)$, as well as the viewing inclination angle $i$, as these are required to fully specify each model \ha\ profile.\footnote{The parameter ranges in Table~\ref{tab:dparams} simply define a {\it computational grid\/} of models, and there is no guarantee every model is realized by an actual star. In addition, some parameter combinations do not result in \ha\ emission and hence fail the requirement for a CAe star.} The viewing inclination angle distribution is straightforward: for randomly-oriented stellar rotation axes, the inclination angle $i$ follows the $p(i)\,di=\sin i\,di$ distribution for the observer \citep{gray2022}. The distributions for $(\rho_0,n)$ are less straightforward, but it reasonable to model these distributions as Gaussians\footnote{Technically the distribution for $\rho_0$ is log-normal as $\log\rho_0$ is taken to be normally distributed.} specified by a mean and standard deviation \citep[see, for example,][]{Sigut2023}. Thus to build random samples of calculated \ha\ profiles, we generate the inclination angle $i$ from the $\sin i$ distribution and generate $(\rho_0,n)$ from
\begin{equation}
\log\rho_0=\mu_\rho + \sigma_\rho\,r_N(0,1)\;,
\end{equation}
and
\begin{equation}
n=\mu_n + \sigma_n\,r_N'(0,1)\;.
\end{equation}
Here $(\mu_\rho,\sigma_\rho)$ are the mean and standard deviation for the $\log\rho_0$ distribution, $(\mu_n,\sigma_n)$ are the mean and standard deviation for the $n$ distribution, and $r_N(0,1)$ and $r_N'(0,1)$ are normally-distributed random numbers with a mean of zero and a standard deviation of one. We take the outer disk disk radius, $R_d$, to be uniformly distributed between $5$ and $50\,R_*$, i.e.\
\begin{equation}
R_d=\left(5+45\,r\right)\,R_*
\end{equation}
where $r$ is a uniform deviate, $0\le r\le1$.

To fit an observed distribution of \ha\ equivalent widths for a specific spectral type, we follow the above procedure to generate a random sample of 100 computed \ha\ profiles for models with a selected viscous $\alpha$ parameter and compare the CDF of the model \ha\ equivalent widths to the observations. The random sample includes only models where the \ha\ profile is classified as having an emission component in order to match the selection procedure for the observed sample (i.e.\ selection for CAe stars).
The comparison is a two-sample KS test with the goodness-of-fit parameter taken as the probability that one would obtain a KS test statistic (maximum different between the CDFs) equal to or larger than observed \citep{Wall2003}. As the inclination angle ($i$) and outer disk radius ($R_d$) distributions are fixed, this comparison is be repeated over the 1512 combinations of the $(\mu_\rho,n_\rho;\mu_n,\sigma_n)$ values listed in Table~\ref{tab:searchgrid} in order to maximize this probability and define the best-fit model. These values are chosen to span the known range of disk density parameters for the CBe stars. It is important to note that our goal is not to extract good measures of these parameters from the data; our real interest is if there is a consistent set of parameters that can reproduce the data for each $\alpha$ viscosity parameter considered.

\begin{table}
\caption{Mean and standard deviation parameters used for the distributions of $\log\rho_0$ and $n$.}
\label{tab:searchgrid}
\begin{center}
\begin{tabular}{ll}\hline\hline
Parameter  &  Values \\\hline
$\mu_\rho$    & (6): $-11.50$, $-11.25$, $-11.00$, $-10.75$, \\ 
              & \hspace{0.56cm} $-10.50$, $-10.25$ \\
$\sigma_\rho$ & (6): $0.10$, $0.20$, $0.40$, $0.60$, $0.80$, $1.00$ \\
$\mu_n$       & (7): $2.00$, $2.25$, $2.50$, $2.75$, $3.00$, $3.25$, \\
              & \hspace{0.55cm} $3.50$ \\
$\sigma_n$    & (6): $0.10$, $0.20$, $0.40$, $0.60$, $0.80$, $1.00$ \\\hline
\end{tabular}
\end{center}
\end{table}

Given this procedure, there are several options as to how to proceed: (1) adopt $(\mu_\rho,n_\rho;\mu_n,\sigma_n)_{Be}$ values derived for the Be stars \citep[use, for example,][]{Sigut2023}; (2) fit the A0 spectral type, for which there is essentially no dependence of the \ha\ distribution on $\alpha$, and then use the obtained best $(\mu_\rho,n_\rho;\mu_n,\sigma_n)_{A0}$ values to generate the samples for the remaining spectral types, A1 through A4, and $\alpha$ values; (3) fit individual $(\mu_\rho,n_\rho;\mu_n,\sigma_n)$ parameters to the observed \ha\ distribution for each spectral type and $\alpha$ value, and then search for consistency in these parameters across the spectral types. 

After some trial and error, we have adopted approach (3). The reasoning here is two-fold: firstly, adopting distributions appropriate for the Be stars, such as those of \citet{Silaj2010a} or \citet{Sigut2023}, would include the influence of many high-mass Be stars which are much different than the cool Ae stars considered here. Even within the CBe stars, there is evidence of different disk parameters between high and low mass CBe stars \citep{Arcos2017}. Secondly, fixing the distribution to that found for the A0 stars (or the Be stars as in option~1) does not work well in practice. The {\it best-fit\/} disk density model for the A0 stars fails to reproduce the \ha\ CDFs for the cooler spectral types. However, this is not an inconsistency because, in practice, a large number of distributions will fit the observed data nearly as well as the best-fit model; therefore, one ends up in searching over a large number of parameter combinations for consistency, and this begins to look like option (3) in the end.

\begin{figure*}
\centering
\includegraphics[width=1.0\textwidth]{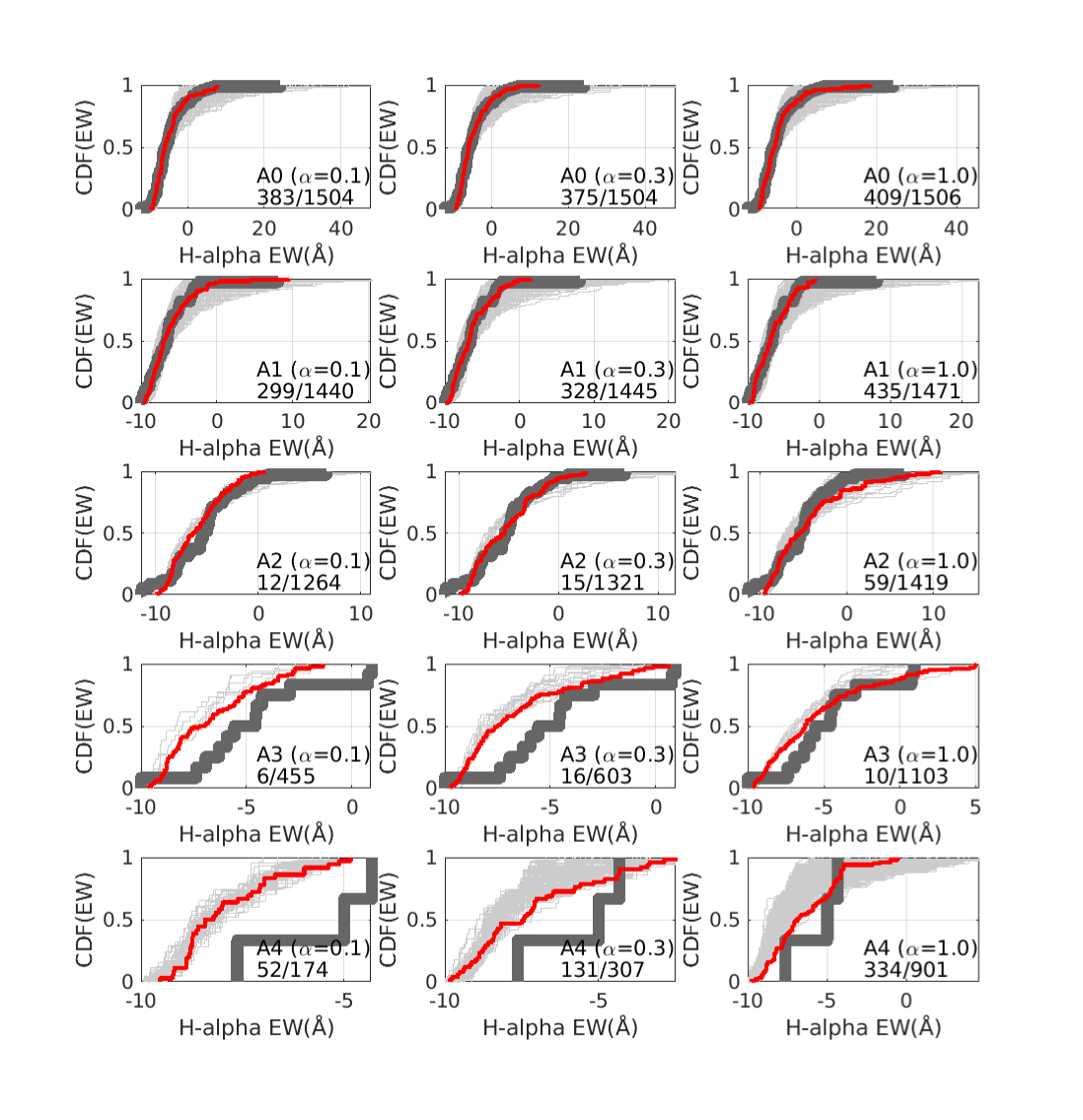}
\caption{Best model fits to the CDF of the observed \ha\ EWs shown in Figure~\ref{fig:ewhist}. The five rows correspond to spectral types A0 through A4 (top to bottom), and the three columns correspond to three $\alpha$ values controlling the amount of shear heating: $\alpha=0.1$ (left), $0.3$ (middle) and $1.0$ (right). The wide, dark gray line in each panel is the observed \ha\ CDF (and is the same across each row). The red line is the \ha\ CDF for the best-fit computational sample, and the light gray lines are all computed samples that satisfy $\log P \ge -0.60$ (5\% level). The number of such samples, out of the total number of samples checked, is also shown in each panel.}
\label{fig:fitCDFs}
\end{figure*}

The result of this procedure is shown in Figure~\ref{fig:fitCDFs}. For each combination of spectral type, A0 through A4, and $\alpha$ parameter, $\alpha=0.10$, $0.30$, and $1.00$, the observed \ha\ EW CDF of Figure~(\ref{fig:ewhist}) is compared to the various model CDFs as defined by $(\mu_\rho,n_\rho;\mu_n,\sigma_n)$. The best-fit model CDF is shown, as well as all models that fit at the 5\% level or better, based on the two-sample KS test. As can be seen from the figure, good fits to the observed \ha\ CDFs can be obtained for the A0 and A1 spectral types, and there is little dependence of these theoretical CDFs on the $\alpha$ parameter used. The best-fit parameters $(\mu_\rho,n_\rho;\mu_n,\sigma_n)$ are discussed below. 

There is a interesting trend in the fits with spectral type shown in Figure~\ref{fig:fitCDFs}. Each panel also gives, for that spectral type and $\alpha$, the number of models out of number tested that fit with a two-sample KS probability of 5\% or higher.\footnote{From Table~\ref{tab:dparams}, there are a maximum of 1512 models, but it provided difficult to build random samples of 100 stars with \ha\ emission for some choices of the parameters $(\mu_\rho,n_\rho;\mu_n,\sigma_n)$, particularly for later spectral types (A3 and A4) with little shear heating (i.e.\ small $\alpha$). Only parameters that could furnish samples of 20 or more (out of 200 attempts) were retained in the comparison.} For spectral types A0 and A1, several hundred of the parameter combinations of Table~\ref{tab:dparams} fit at least at the 5\% level, with the $\alpha=1.00$ models having slightly more fits. However, this changes for spectral types A2 and A3. The number of fitting models is now down to the single digits for $\alpha\le0.3$, whereas tens of models fit for $\alpha=1.00$. Recall that A2 is where shear heating first becomes important. At A4, the number of fits rises again, with the most for $\alpha=1.0$, but this is really a consequence of having an observational sample size of just three. 

\begin{figure*}
\centering
\includegraphics[width=1.0\textwidth]{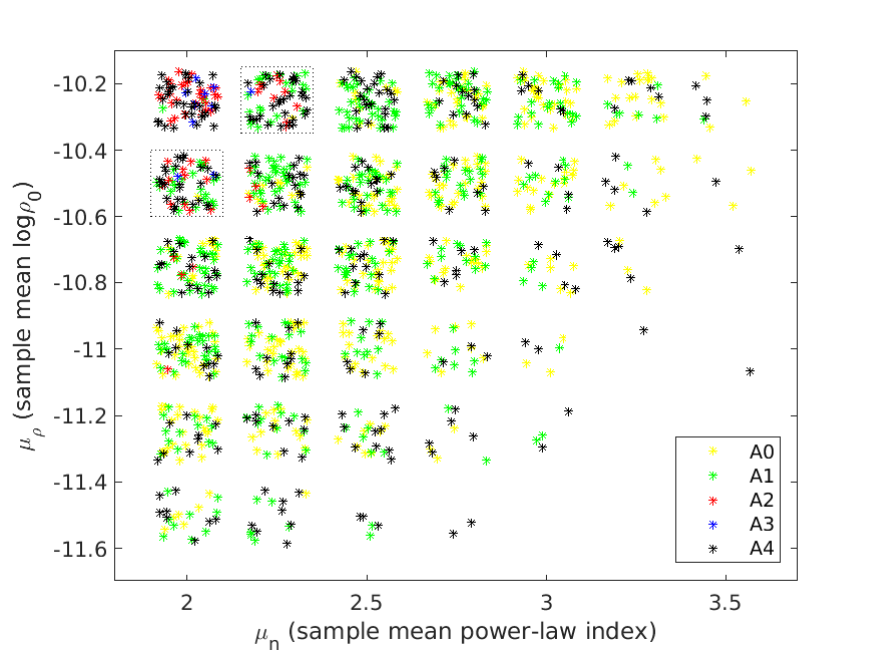}
\caption{Parameters $(\mu_\rho,\mu_n)$ for the theoretical \ha\ CDFs fits of Figure~\ref{fig:fitCDFs} corresponding to the rightmost column of Figure~\ref{fig:fitCDFs} for $\alpha=1.0$. All models have KS probabilities $\log(P)>\log(0.05)$. The symbols are color-coded by spectral type as identified in the legend. Corresponding to each $(\mu_\rho,\mu_n)$ are a number of CDFs corresponding to different choices for $(\sigma_\rho,\sigma_n)$ (see Table~\ref{tab:searchgrid}). These models are randomly jittered about their common $(\mu_\rho,\mu_n)$ values for clarity. The two dotted squares in the upper left corner mark consistent samples for which all five spectral types occur for the same $(\mu_\rho,\mu_n)$.}
\label{fig:paramCDFs}
\end{figure*}

What about the consistency of these fits, i.e.\ can all of the CAe \ha\ observations be modeled with a single choice for $\mu_\rho$ and $\mu_n$, i.e.\ a single choice for the distributions of disk density parameters? This is addressed in Figure~\ref{fig:paramCDFs} which plots, for the $\alpha=1.00$, the model samples that fit to 5\% or better in the $(\mu_\rho,\mu_n)$ plane. Here a point is placed at the point $(\mu_\rho,\mu_n)$ if the \ha\ EW CDF of a model sample with these parameters fits the observed CDF to 5\% or better (color-coded by spectral type). There can be many points at a single $(\mu_\rho,\mu_n)$ because there are many choices for the additional $(\sigma_\rho,\sigma_n)$ parameters; these are jittered for clarity.  As can be seen from the figure, wide portions of the parameter space are consistent with the A0 and A1 \ha\ strengths, due to the intrinsically stronger \ha\ emission, and the A4 \ha\ strengths, due to the small observed sample not providing a strong constraint. The real test is the intermediate A2 and A3 spectral types where the \ha\ strengths are rapidly falling, the influence of shear heating is increasing, and the sample sizes are large enough to provide meaningful constraints. For $\alpha=1.0$, there is a small portion of the parameter space spanned by Table~\ref{tab:dparams} where all spectral types, A0 through A4, can be fit, and this region occurs in the upper left corner of the figure where $(\mu_\rho,\mu_n)\sim (-10.50,2.00)$. For the other $\alpha$ values, $\alpha\le0.3$ (not shown in the figure), no such common region can be found. 

We do not want to over-interpret this result. This method is not a sensitive way to determine the underlying $(\log\rho_0,n)$ distributions for a sample of stars, as clearly demonstrated by the large number of fits at spectral types A0 and A1. However, there is a hint in the analysis that the strengths of the \ha\ line, particularly for A3 and A4 (where shear heating has been theoretically demonstrated to be important) are not consistent with $\alpha\le\approx\!0.3$. This is, of course, contrary to the previous discussion focused on CAe star counts, and apparently inconsistent with the absence of CAe stars of spectral type A5 or later (see Figure~\ref{fig:TdiskTeff}). More sensitive searches for weak \ha\ emission at spectral types A3 and later would be very helpful in clarifying this situation.

\section{Conclusions and Future Work}
\label{sec:Conclusions}

We have constructed theoretical models for the temperature structure of circumstellar disks surrounding A-type main sequence stars that include both radiative heating from the central central star and viscous shear heating from the disk's Keplerian rotation, with the viscosity modeled by the $\alpha$ prescription of \cite{Shakura1973a}. For spectral types A0 and A1, the contribution of shear heating, even with $\alpha=1.0$, is small and makes no significant impact on the predicted disk temperatures or \ha\ emission line strengths. However, for spectral type A2 and cooler, shear heating can become significant, and the choice of $\alpha$ affects the strength of the \ha\ emission. In particular, shear heating with $\alpha\ge\sim 0.3$ can prevent CAe star disks from becoming as cool as predicted by the trend for average disk temperatures of $\sim 0.6\,T_{\rm eff}$ based on radiative heating alone. All models show a significant decline in the number of CAe stars near spectral type A2 based on calculated \ha\ emission; however, this decline is moderated by increasing $\alpha$. Models with $\alpha=1.0$ also predict significant emission for spectral type A5, although no such objects have been observed. 

The dependence of the presence and strength of \ha\ emission on shear heating for latter spectral types makes it possible to constrain the disk viscosity $\alpha$ parameter using observed samples of CAe stars, both in terms of number of classified CAe stars as a function of spectral type and on the strength and distribution of the \ha\ emission equivalent width. This radiative method is a novel approach for CAe stars as previous measurements of $\alpha$ in CBe circumstellar disks have relied on hydrodynamical modeling with the implied viscous timescale, $t\sim R^2/\nu$, to match observed variations of disk emission. However, this radiative method is limited to providing some sort of ``ensemble average" of $\alpha$ over the CAe population, as opposed to being extracted for individual objects. If the underlying physical process for the viscosity is, for example, magnetic fields via the magneto-rotational instability \citep{Balbus1991}, variations in the magnetic field strength and/or geometry might give rise to variations in $\alpha$ between objects, something this radiative approach cannot capture. Nevertheless, given the central role viscosity plays in the circumstellar disks of CAe and CBe stars, all constraints are valuable.

We have attempted to extract such an ensemble-averaged $\alpha$ for the CAe sample of \citet{Anusha2021}. The results are not consistent. The drop in the fraction of CAe stars as a function of spectral type in the range A0 through A4 seems most consistent with $\alpha\le\,\approx\!0.1$, although there are large uncertainties as the number of A3 and A4 stars in the sample is small. On the other hand, modeling the observed \ha\ emission line strength as a function of spectral type through the CDF of the \ha\ equivalent width suggests that $\alpha$ is larger, with the evidence coming from spectral types A3 and A4; however, the number of detected objects for these later spectral types is small. In addition, $\alpha=1.0$ models lead to the expectation of spectral type A5 CAe stars (see Figure~\ref{fig:TdiskTeff}) yet none are seen. This disagreement in $\alpha$ for CAe counts versus \ha\ EW CDF is likely due to the statistics of small numbers, and steady improvements will be realized as the number of CAe stars, particularly for latest spectral types, increases.

As a follow-up to this study, we plan to match all the available \ha\ spectra for the 159 CAe stars in the \citet{Anusha2021} sample using the \texttt{Bedisk}/\texttt{Beray} code suite to obtain best-fit disk density parameters $(\rho_0,n,R_d)$ and inclination angles for each individual star. This will allow a detailed comparison with the known distributions for the CBe stars of \citet{Silaj2010a} and \citet{Sigut2023}.

A final point we have not addressed in this work is the potential role of A shell-stars, where \ha\ emission is not detected but deep central absorption well below the expected photospheric profile is seen. Such objects are a natural extension of the CAe stars in which the inclination of the system is high, $i\ge\approx\! 80^\circ$, in which the observer's line of sight passes directly through the thin, equatorial disk. \ha\ shell stars are known to extend to later spectral types than the CAe stars, predominantly seen to spectral type A5, with a decline towards A7 and near disappearance after that \citep{Slettebak1982, Hanuschik1996, Hauck2000}. The \ha\ calculations in this work also naturally predict the frequency and absorption strength of the \ha\ shell feature as a function of spectral type and we will examine this line of reasoning in a subsequent work.

\begin{acknowledgments}
The authors are grateful to the anonymous reviewer for their constructive feedback and careful reading of the manuscript, which helped enhance both the clarity and quality of our work.
TAAS gratefully acknowledges funding from the Natural Sciences and Engineering Research Council of Canada (NSERC) through the Discovery Grant program. 
\end{acknowledgments}

\clearpage

\newpage

\bibliography{main}{}
\bibliographystyle{aasjournal}

\end{document}